\documentclass[12pt,a4paper,preprintnumbers,amsmath,amsfonts,superscriptaddress,nofootinbib]{revtex4}
\usepackage[dvips]{color}
\usepackage[dvipdfm]{hyperref}
\usepackage{multirow,slashbox}
\usepackage{graphics}

\begin{document} 

\title{Viability of $f(R)$ Theories with Additional Powers of Curvature}
\author{Anthony W. Brookfield}
\email[Email address: ]{php04awb@sheffield.ac.uk}
\affiliation{Astroparticle Theory and Cosmology Group, Department of Applied Mathematics, The University of Sheffield, Hounsfield Road, Sheffield S3 7RH, United Kingdom}
\affiliation{Astroparticle Theory and Cosmology Group, Department of Physics and Astronomy, The University of Sheffield, Hounsfield Road, Sheffield S3 7RH, United Kingdom}
\author{Carsten van de Bruck}
\email[Email address: ]{C.vandeBruck@sheffield.ac.uk}
\affiliation{Astroparticle Theory and Cosmology Group, Department of Applied Mathematics, The University of Sheffield, Hounsfield Road, Sheffield S3 7RH, United Kingdom}
\author{Lisa M.H. Hall}
\email[Email address: ]{Lisa.Hall@sheffield.ac.uk}
\affiliation{Astroparticle Theory and Cosmology Group, Department of Applied Mathematics, The University of Sheffield, Hounsfield Road, Sheffield S3 7RH, United Kingdom}
 
\begin{abstract} 
We consider a modified gravity theory, $f(R)=R-a/R^n+bR^m$, in the metric formulation, which has been suggested to produce 
late time acceleration in the Universe, whilst satisfying local fifth-force constraints.  We investigate 
the parameter range for this theory, considering the regimes of early and late-time acceleration, 
Big Bang Nucleosynthesis and fifth-force constraints. We conclude that it is difficult to find a unique 
range of parameters for consistency of this theory.
\end{abstract}

\maketitle

\section{Introduction}
Various cosmological observations suggest that the universe is pervaded by a 
new form of energy, dubbed dark energy, giving rise to accelerated expansion 
at the present epoch \cite{Riess:1998cb,Perlmutter:1998np,Tonry:2003zg}. These observations are a challenge for 
fundamental physics, since a well--motivated candidate for dark energy has to be found. The usual 
candidates include scalar fields or extra dimensions.
 
Recently, it has been suggested that instead of a new matter form, the cosmic accelerated expansion 
could be attributed to a modification of General Relativity itself, see e.g. 
\cite{Deffayet:2001pu,Freese:2002sq,Dvali:2003rk,Nojiri:2003ni,Carroll:2003wy,Capozziello:2003tk,
Nojiri:2003ft,Carroll:2004de,Abdalla:2004sw}. 
For reviews see e.g. \cite{Easson:2004fq,Nojiri:2006ri,Trodden:2006qk}. 
The simplest models of modified gravity are theories 
in which the Ricci scalar $R$ in the Einstein--Hilbert Lagrangian is replaced by some (arbitrary) 
function of $R$. Examples of these models include inverse powers $f(R) = R^{-n}$ or 
$f(R) = R - a R^{-n}$. It has been shown that these models are equivalent to scalar tensor theories 
with vanishing Brans--Dicke parameter, thereby ruling out these models \cite{Chiba:2003ir}. 
It has recently been argued that by adding terms like $bR^m$ ($m$ positive), the 
field can be made massive and hence short-ranged, allowing the theory to be 
made consistent with local constraints~\cite{Nojiri:2003ft}.  However, past work considered 
only the vacuum of the theory, without taking into account the coupling to matter. 

In this paper, we will consider theories with positive and negative curvature 
terms, taking the matter coupling into account. This is a non-trivial task and we have to make 
simplifying assumptions in order to proceed. The paper is organised as follows. In the next Section 
we formulate the theory both in the Jordan frame and in the Einstein frame. We will write down 
useful equations needed for later Sections. In Section \ref{seclocal}, local constraints for the 
choice $f(R)=R+bR^m$ are considered, analytically and numerically for $m=2$ and numerically for 
more general cases. In Section \ref{secDE} we discuss consequences for the model $f(R)=R-aR^{-n} + bR^m$. 
We will summarise our findings in Section \ref{seccomp}.

\section{Jordan-- and Einstein Frame Formulation}
\label{sectheory}
The theory we are going to consider is specified by the action
\begin{equation}
{\cal S}_{\rm JF} = \int d^4 x \sqrt{-g}\left(\frac{1}{\kappa^2}f(R) + {\cal L}_{\rm m}\right),
\end{equation}
where $R$ is the Ricci scalar and ${\cal L_{\rm m}}$ is the matter Lagrangian. 
From this action, the field equations can be easily derived:
\begin{equation}\label{fieldequation}
R_{\mu\nu} f'(R) - \frac{1}{2}g_{\mu\nu}f(R) + g_{\mu\nu}\Box f'(R) - 
\nabla_\mu\nabla_\nu f'(R) = \frac{\kappa^2}{2}T_{\mu\nu},
\end{equation}
where $f' = df/dR$. The trace of this equation gives
\begin{equation}\label{Requation}
\Box R + \frac{f^{'''}(R)}{f^{''}(R)}\nabla_\lambda R \nabla^\lambda R 
+ \frac{Rf'(R)}{3f^{''}(R)} - \frac{2f(R)}{3f^{''}(R)} = \frac{\kappa^2}{6f^{''}(R)}T.
\end{equation}

Spherically symmetric solutions of $f(R)$ theories have been recently studied in 
\cite{Brevik:2004sd,Cognola:2005de,Multamaki:2006zb}. In particular 
it was found that the Schwarzschild--de Sitter metric is an exact solution to the field equations for a large 
class of models. 

%***Define metric conventions (-,+,+,+) and quote SdS paper - comment on expanding locally around Minkowski
%
%\begin{equation}
%\frac{d^2R}{dr^2}+\frac{2}{r}\frac{dR}{dr}+\frac{f'''}{f''}\left(\frac{dR}{dr}\right)^2 - \frac{dV_{eff}^{(R)}}{dR} =0.
%\label{fullR}
%\end{equation}

Furthermore, it is well known that the theory can be rewritten as a scalar tensor theory (e.g. \cite{Flanagan:2003iw,Nojiri:2003ft}). 
To see this, introduce two auxiliary fields, $A$ and $B$, and write the gravitational part of the full 
action as 
\begin{equation}
{\cal S}_{\rm JF} = \frac{1}{\kappa^2}\int d^4x \sqrt{-g} \left(B(R-A)+f(A)\right).
\end{equation}
Making the variation with respect to $B$ gives $A=R$, whereas the variation with respect to $A$ gives 
\begin{equation}
B=f'(A),
\end{equation}
allowing us to eliminate $B$ and write the gravitational part of the action as 
\begin{equation}\label{reform}
{\cal S}_{\rm JF} = \frac{1}{\kappa^2}\int d^4x \sqrt{-g} \left(f'(A)(R-A)+f(A)\right).
\end{equation}
Variation with respect to $A$ gives 
\begin{equation}\label{aquation}
f''(A)(R-A) = 0.
\end{equation}
This equation is solved by $R=A$ if $f''$ is non-vanishing.

%We will assume that 
%$R=A$ is also true at points for which $f''$ vanishes and for the reminder of the 
%paper we will not distinguish between $A$ and $R$.  

To obtain a Einstein frame formulation of the theory, perform a conformal 
transformation $\tilde{g}_{\mu\nu} = e^\sigma g_{\mu\nu}$ of the action (\ref{reform}) 
and choose 
\begin{equation}\label{sigmadef}
\sigma = -\ln f'(A).
\end{equation}
The gravitational sector in the Einstein frame then reads (Einstein frame quantities are 
denoted with a tilde) 
\begin{equation}
{\cal S}_{\rm EF} = \frac{1}{\kappa^2}\int d^4x \sqrt{-\tilde{g}}\left(\tilde{R} 
- \frac{3}{2}\left(\tilde{\nabla} \sigma \right)^2 - V(\sigma)\right),
\end{equation}
with 
\begin{equation}
V(\sigma) = \frac{A}{f'(A)} - \frac{f(A)}{f'(A)^2},   
\label{eqnV}
\end{equation}
where the relation between the field $A$ and 
$\sigma$ is given by Eqn (\ref{sigmadef}). 

A few comments are in order. Firstly, for the conformal transformation to exist, $f'$ has to be 
non--zero and positive, as it can be seen from Eqn (\ref{sigmadef}). We will always consider an $f(R)$ 
such that this is the case. Secondly, in order to identify the field $A$ with the {\it Jordan frame} 
Ricci scalar $R$, $f''$ must be non--zero, otherwise the solution to Eqn (\ref{aquation}) is not unique. 
However, we will assume that even at the (isolated) points in which $f''$ vanishes, $A=R$ is true.  We will not distinguish between $A$ and $R$ for the remainder of this paper.

From these considerations it has been shown that the theory is equivalent to a scalar--tensor theory 
with vanishing Brans--Dicke parameter $\omega_{\rm BD}$. This is in contrast 
with current observations, which demand $\omega_{\rm BD}>40000$, if the field 
is long--ranged in the solar system. However, for 
\begin{equation}\label{ourtheory}
f(R) = R - \frac{a}{R^n} + bR^m
\end{equation}
with $a$ and $b$ positive and non--zero, it has been argued that the scalar degree of freedom 
$\sigma$ can be made massive, leading to a short--ranged force and thereby 
avoiding the conflict with local (i.e. solar--system) experiments \cite{Nojiri:2003ft}.

%DO WE NEED THE FOLLOWING??? Considering this theory, the condition that $f''(R)\ne 0$ leads to
%\begin{eqnarray}
%R^{n+m}\ne \frac{n(n+1)a}{m(m-1)b}.
%\label{f''valid}
%\end{eqnarray}

Considering a flat Robertson--Walker universe filled with dust, 
the Einstein--frame field equations give 
\begin{equation}
{\tilde H}^2 = \frac{\kappa^2}{6}\left(\tilde{\rho}_{\rm m} + \rho_{\rm \sigma}\right)
\end{equation}
and
\begin{equation}
\ddot\sigma + 3\tilde{H}\dot\sigma + \frac{1}{3}\frac{\partial V}{\partial \sigma} = 
-\frac{\beta\kappa^2}{3}\tilde{\rho}_{\rm m}.
\end{equation}
In the last equation, $\beta$ specifies the coupling between matter and the field $\sigma$ and 
is given by $\beta=1/2$. The relation between $\beta$ and the Brans--Dicke parameter 
$\omega_{\rm BD}$ is given by $\beta^2 = 3/(2\omega_{\rm BD}+12)$, so that $\beta=1/2$ 
corresponds to $\omega_{\rm BD}=0$.
The energy conservation equation for matter is given by
\begin{equation}
\dot{\tilde{\rho}}_{\rm m} + 3 \tilde{H}\tilde{\rho}_{\rm m} 
= 2\beta\dot\sigma \tilde{\rho}_{\rm m}.
\end{equation}

Since the theory was introduced to obtain an accelerating universe at the present epoch 
without to resort to an additional energy form, we fix 
\begin{equation}\label{avalue}
a = (10^{-42} {\rm GeV})^{2(n+1)}, 
\end{equation}
following \cite{Nojiri:2003ft} and \cite{Carroll:2004de}. Note that, with $a\neq0$, Minkowski space 
is not a solution of the field equations. 

\subsection{Properties of the Potential}
In this paper we are interested in the theory given by Eqn (\ref{ourtheory}) for the function $f(R)$, for which 
the potentials in the Jordan and Einstein frame have interesting properties. The effective potential 
for $R$ (see Eqn~(\ref{Requation})), fulfils 
\begin{equation}
\frac{dV_{\rm eff}^{(R)}}{dR} = \frac{2f(R) -Rf'(R) - \beta\kappa^2\rho}{3f''(R)}
\end{equation}
where we have assumed a pressureless fluid with $T=-\rho$.

The extrema of the potential are given by $R$, such that
\begin{equation}\label{effectivepotential}
2f(R) - R f'(R) - \beta\kappa^2 \rho = 0.
\end{equation}
For gravitational stability (i.e. no instability for $R$), we require that the extremum is a minimum,
\begin{equation}
\frac{d^2V}{dR^2}\Bigr\rvert_{\rm min} >0
\end{equation}
and using Eqn (\ref{eqnV}), we find
\begin{equation}
0<f'' < \frac{f'}{R}.
\label{stability}
\end{equation}

In regions of high curvature, such as during Big Bang Nucleosynthesis (BBN) or locally on Earth, 
we expect to be able to ignore the $a$ term. In this regime, the minimum is specified by
\begin{equation}
R_{\rm min} + (2-m)bR^m_{\rm min} - \beta\kappa^2\rho = 0. 
\label{minexist}
\end{equation}
Since $\rho$ is positive, a solution with positive $R$ exists only when $m<2$ or the magnitude of the second term is smaller 
than $R_{\rm min}$. In the latter case we find that a good approximation for the minimum is given by 
$R_{\rm min}\approx \beta\kappa^2 \rho$. This, however, is not true for $m<2$, but holds in general for $R>bR^m$. 

The effective mass at the minimum ($R=R_{\rm min}$) is given by
\begin{eqnarray}
\mu^2 &=& \frac{d^2 V_{\rm eff}^{(R)}}{dR^2} = \frac{f'(R_{\rm min}) 
- f''(R_{\rm min})R_{\rm min}}{3f''(R_{\rm min})} \nonumber \\
&=&\frac{(3-m)R^{2-m}}{3m(m-1)b} - \frac{2(m-2)R}{3(m-1)m} 
- \frac{\beta\kappa^2\rho(2-m)R^{m-1}}{3(m-1)mb} \nonumber \\
&\approx& \frac{R^{2-m}_{\rm min}}{3(m-1)mb}.
\label{eqnmass}
\end{eqnarray}
In the last two lines we have ignored the $aR^{-n}$--term. Note that the mass is dependent on 
the ambient matter density. 

\section{Local Constraints for $f(R)=R+bR^m$}
\label{seclocal}
In regions of high curvature, for example locally on Earth, one might expect the inverse curvature terms 
to be sub-dominant. In this section, we will study this regime of the theory and investigate the effect 
of an additional curvature term resulting in a local fifth force and set $a=0$.  We note, however, that 
if we require only small corrections to Einstein gravity, we may assume that
\begin{equation}
b< R^{1-m}
\label{bupperlim}
\end{equation}
to an order of magnitude estimate.  This limit is consistent with the existence of a minimum, 
Eqn (\ref{minexist}). For a range of $b$, we discuss the strength and range of the resultant force, in 
order to test if this is a viable alternative theory of gravity. In doing so we will solve Eqn (\ref{Requation}) 
and not the full field equation (\ref{fieldequation}). Although limiting, this method is easier and 
provides insights into the predictions of the theory, taking matter couplings into account. 
It also allows us to study extended objects (like the Earth), instead of point particles. 

Fifth-force experiments have placed strong constraints on the strength of any deviation of the gravitational potential from that predicted by Einstein's theory of General Relativity.  
It is usual to assume a Yukawa potential for the fifth force,
\begin{eqnarray}
V_{\rm Yuk}(r)=-\alpha \frac{M_1M_2e^{-r/\lambda}}{8\pi M_{Pl}^2r}
\label{yuk}
\end{eqnarray}
where $\alpha$ is a measure of the strength of the force relative to gravity, and $\lambda$ describes 
the range over which the force acts. The allowed strength of such a force is constrained for a large 
range of scales from $10^{-6} - 10^{14}$m \cite{Adelberger:2003zx}. If this theory of modified gravity 
is to prove viable, it must satisfy the experimental constraints on all scales.  

For simplicity, we consider the simple system of the Earth sitting in the local solar system medium.  
We assume that the Earth has a radius of $r_E=6.7\times 10^9$mm and a constant density of $5.5{\rm g/cm}^3$ 
(which corresponds to approximately $\kappa^2\rho_E\approx 10^{-53}{\rm GeV}^2$). 
The inter-planetary medium has an approximate density of $10^{-24}{\rm g/cm}^3$ ($\kappa^2\rho_{SS}\approx 10^{-78} {\rm GeV}^2$), which we also assume to be constant.  We neglect the effect of the other solar-system bodies (in particular the sun), and the Earth's atmosphere (which we will justify later).

In this simple system, it is possible to consider the force on satellites in orbit around the Earth. 
Geostationary satellites sit at approximately $6r_E$, corresponding to around $10^7$m.  
At this level, the strength of a fifth force is constrained to $\alpha<10^{-8}$.

If we consider a time-independent, radially symmetric solution, Eqn (\ref{Requation}) becomes
\begin{equation}
\frac{d^2R}{dr^2}+\frac{2}{r}\frac{dR}{dr}+\frac{f'''}{f''}\left(\frac{dR}{dr}\right)^2 
- \frac{dV_{\rm eff}^{(R)}}{dR} =0.
\label{fullR}
\end{equation}

We shall only consider a subset of the theory and consider models in which $m=\frac32,2,\frac52,3$.  

\subsection{$m=2$}
\label{secm=2}

We begin by considering a specific case of our theory, where $f(R)$ takes the form
\begin{equation}
f(R)=R+bR^2.
\end{equation}
To check our results, we will consider solutions to both the Jordan frame $R$ equation, 
and the equivalent Einstein frame equation for $\sigma$. The range of $b$ (as an order of 
magnitude) is given by
\begin{eqnarray*}
b<10^{53}{\rm GeV}^{-2}, 
\end{eqnarray*}
as can be seen from Eqn (\ref{bupperlim}).

\subsubsection{Jordan Frame $R(r)$}
\label{jordanm=2}
In this simple case, the equation of motion for $R(r)$, given by Eqn (\ref{fullR}) becomes
\begin{eqnarray}
%R^{\prime\prime} + \frac{2}{r}R^{\prime} - \mu^2 R = - \mu^2\beta\kappa^2\rho
\frac{d^2R}{dr^2} + \frac{2}{r}\frac{dR}{dr} - \mu^2 R = - \mu^2\beta\kappa^2\rho
\label{simpleR}
\end{eqnarray}
where $\mu=1/\sqrt{6b}$.  
We assume that $R(r=\infty) = R_{\rm min}$ with $R$ taking the value which minimises the effective potential, $R_{\rm min}=\beta\kappa^2\rho$.  $R_{0}$ and $R_{\infty}$ denote the minima in the test mass and background field respectively.

A solution to Eqn (\ref{simpleR}) is given by
\begin{eqnarray}
R=A\frac{\cosh{\mu r}}{r}+B\frac{\sinh{\mu r}}{r} + R_{\rm min}.
\label{Rgeneral}
\end{eqnarray}
We consider two, solutions, exterior and interior to the test mass.  In order to satisfy our equation, 
we require that $\frac{dR}{dr}=0$ at $r=0$.  The interior solution therefore takes the form
\begin{eqnarray}
R_{\rm int}=\frac{\left(R_i-R_{0}\right)\sinh{\mu r}}{\mu r} + R_{0}
\end{eqnarray}
where $R_i$ is the value of the field at $r=0$.

Externally, we impose two boundary conditions: $R(r=r_c)=R_c$ and $R(r=\infty)=R_{\infty}$, and the external solution 
becomes
\begin{eqnarray}
R_{\rm ext}=\frac{\left(R_c-R_{\infty}\right)r_c e^{-\mu (r-r_c)}}{r} + R_{\infty}
\end{eqnarray}

At the boundary between the test mass and the background, the two solutions and their derivatives must match.  This condition yields the following expressions for $R_i$ and  $R_c$:
\begin{eqnarray}\label{eqn:consts}
R_i&=&\left(R_{\infty}-R_{0}\right)\left(1+\mu r_c\right)e^{-\mu r_c} + R_{0} \\
R_c&=&\frac{\left(R_{\infty}+R_{0}\right)}{2} + \frac{\left(R_{\infty}-R_{0}\right)}{2\mu r_c}\left[1-\left(1+\mu r_c\right)e^{-2\mu r_c}\right]
\end{eqnarray}

\begin{figure}[t] 
\begin{center}
\scalebox{0.5}{\includegraphics{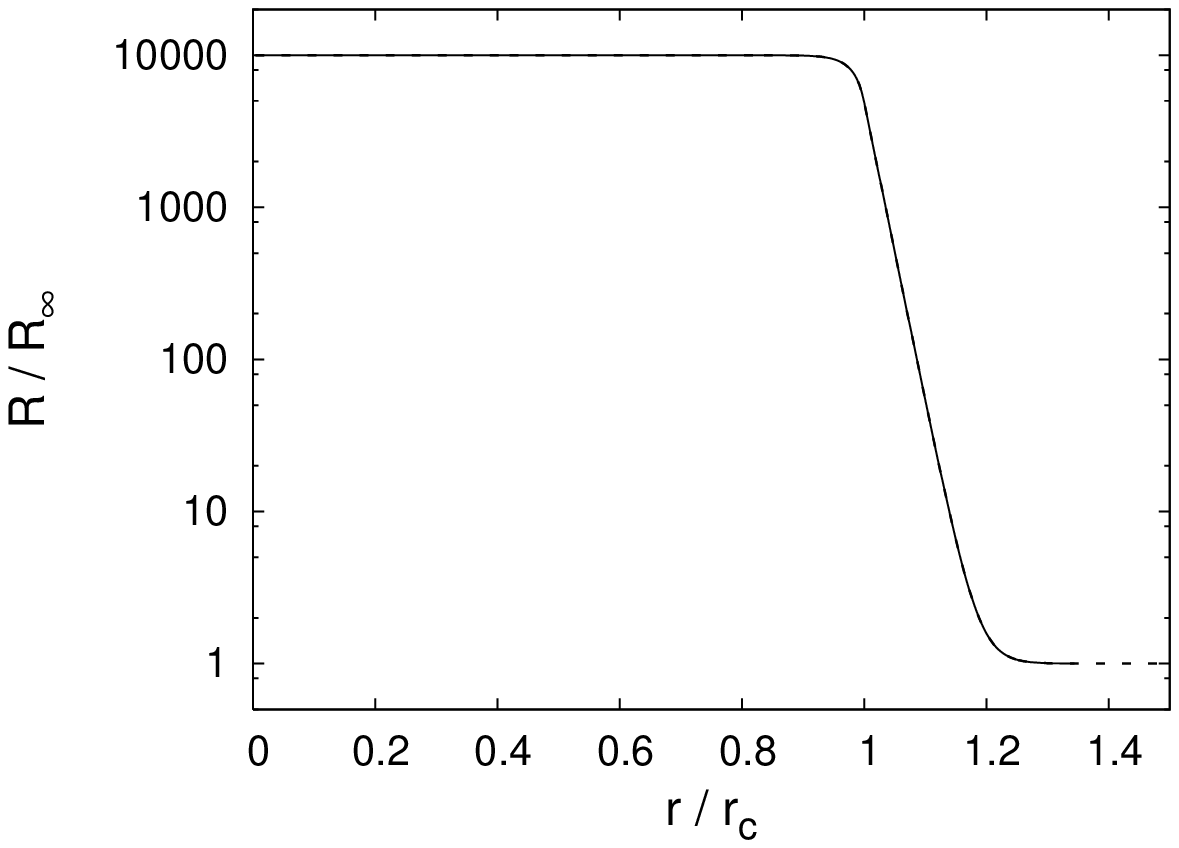}}
\scalebox{0.5}{\includegraphics{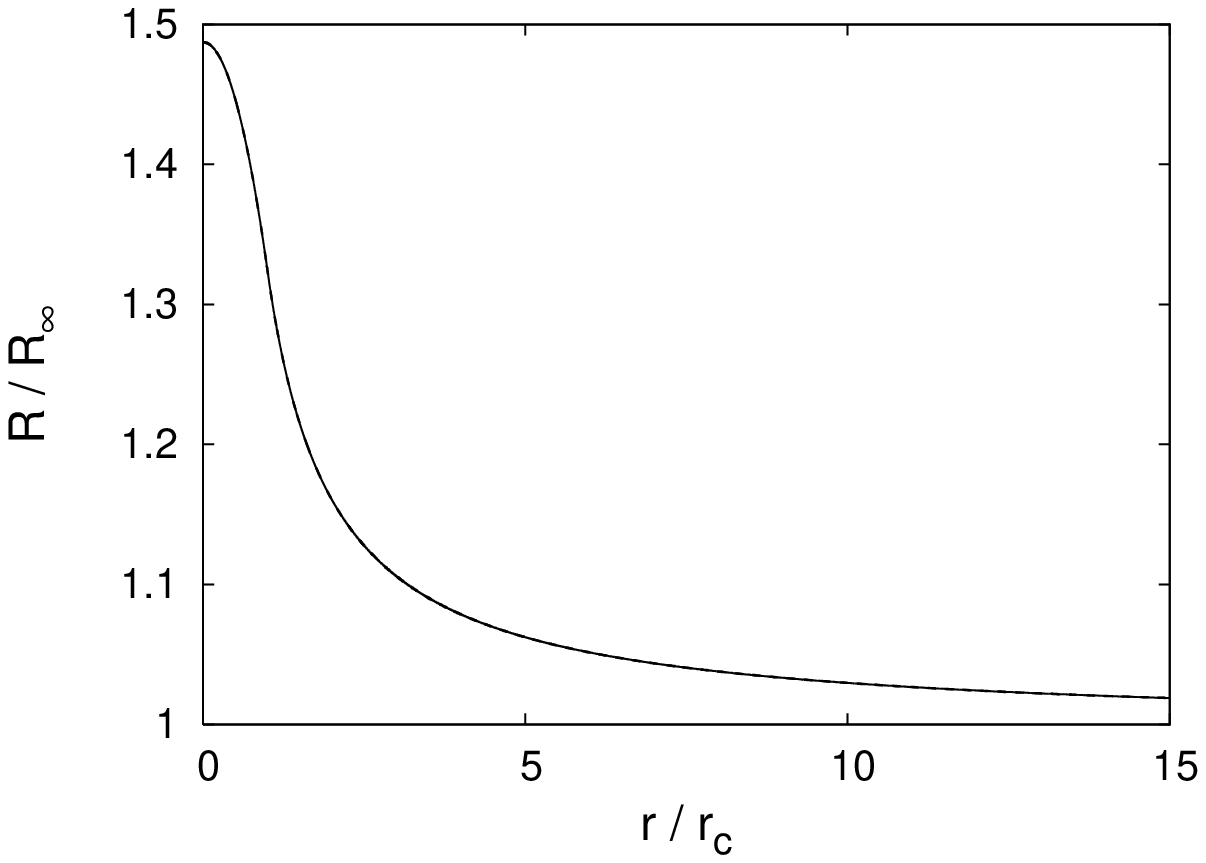}} 
\caption{\label{thickthin}Numerical solutions showing the evolution of $R(r)$ for two choices of $r_c$.  In the left-hand plot, $r_c$ is $10^4$ bigger than in the right-hand plot, so that $\delta_R\rightarrow1$ and $\delta_R\rightarrow0$ respectively. All other parameters are kept constant.  The analytic solutions are shown with dotted lines and exactly follow the numerics.}
\end{center} 
\end{figure}

Given this form of a solution for $R$, we can see that there are two different regimes of behaviour.  
When $\mu r_c>>1$, $R_i\approx R_c$, and the field sits at its minimum inside the body.  As the field approaches 
the boundary into the outside medium it begins to evolve very quickly, and soon settles into the exterior minimum.  
When $\mu r_c\ll 1$, $R_i\approx R_\infty$ and the field remains at a value close to the external minimum.
Two such cases have been modelled numerically and compared to the analytics above, see Figure (\ref{thickthin}).
The reader may note the similarity of these solutions to the thick and thin shelled regimes described in the chameleon model~\cite{Khoury:2003rn}.  We define a parameter, $\delta$ to differentiate between thick ($\delta\rightarrow 0$) and thin ($\delta\rightarrow 1$) regimes:
\begin{equation}
\delta_R = \frac{R_0-R_i}{R_0}.
\label{deltar}
\end{equation}

\subsubsection{Einstein Frame $\sigma(r)$}
Rather than formulating the theory in the Jordan frame, we can consider the equations for the scalar field $\sigma(r)$ in the Einstein frame.  We will see that this has some significant advantages later on.

The equation governing the evolution of $\sigma$ in the Einstein frame is given by
\begin{equation}
\frac{d^2\sigma}{dr^2}+\frac{2}{r}\frac{d\sigma}{dr}- \frac13\frac{dV_{\rm eff}^{(\sigma)}}{d\sigma}=0
\label{fullsigma}
\end{equation}
where
\begin{equation}
\frac{dV_{\rm eff}^{(\sigma)}}{d\sigma} = \frac{Rf'-2f+\beta\kappa^2\rho}{(f')^2}
= \frac{-R+\beta\kappa^2\rho}{(1+2bR)^2}
\end{equation}
This assumes that the conformal transformation relates $R$ to $\sigma$ by
\begin{eqnarray}
e^{-\sigma}=f^{\prime}(R)=1+2bR.
\end{eqnarray}

We have already stated that, in order to study only small deviations to Einstein gravity, we require $bR<1$.  
It is therefore appropriate to assume $f'$ is close to unity and we can expand $e^{-\sigma} \approx 1-\sigma$. 
Hence
\begin{eqnarray}\label{approxi}
\sigma\approx -2bR.
\label{approxsigma}
\end{eqnarray}
In this regime, Eqn (\ref{fullsigma}) becomes 
\begin{equation}
\frac{d^2\sigma}{dr^2}+\frac{2}{r}\frac{d\sigma}{dr}- \frac{\sigma}{6b}-\frac{\beta\kappa^2\rho}{3}=0.
\label{m=2sigma}
\end{equation}
We again require that $\frac{d\sigma}{dr}=0$ at $r=0$, and that $\sigma$ sits in the minimum of its effective potential far away from the test body.  The solution to this system is equivalent to Eqn (\ref{Rgeneral}), taking $\sigma=-2bR$, 
as it can also be seen from using Eqn (\ref{approxi}) in Eqn (\ref{m=2sigma}) to get (\ref{simpleR}).
Similar to Eqn (\ref{deltar}), we can define:
\begin{equation}
\delta_\sigma = \frac{\sigma_0-\sigma_i}{\sigma_0}.
\end{equation}

\subsubsection{Fifth Force}
\label{forcem=2}

In order to calculate the modified gravitational force in the physical frame, we begin by considering the force generated by the equivalent scalar field in the Einstein frame.  

We can calculate the strength of force which would be mediated by $\sigma$ in the Einstein frame,

\begin{eqnarray}
F_E&=&-\beta M\nabla_r\sigma \\
V_E&=&\beta M \sigma. \label{V_E}
\end{eqnarray}
The force measured in the Jordan frame is related to the Einstein frame force via the conformal transformation
\begin{equation}
F_J=e^{-\sigma}F_E=f' F_E
\end{equation}
Note once again that the assumption of perturbed Einstein gravity results in $f'\approx 1$, and 
therefore
\begin{equation}
F_J \approx F_E,
\end{equation}
ignoring higher order corrections. 

The external solution for $R$ is given by,
\begin{eqnarray}\label{eq:Rext}
R_{\rm ext}=\frac{\left(R_c-R_{\infty}\right)r_c e^{-\mu (r-r_c)}}{r}+R_{\infty},
\end{eqnarray}
and so using Eqn (\ref{approxsigma})
\begin{eqnarray}
\sigma(r)\approx -2b\left(R_{c}-R_{\infty}\right)r_c \frac{e^{-\mu (r-r_c)}}{r}+\sigma_\infty.
\end{eqnarray}
Note this is the same equation we might have obtained directly from Eqn (\ref{m=2sigma}).

Ignoring any additional constants, the potential of the field can be immediately seen to be
\begin{eqnarray}
V(r)=-12\beta^2 b\frac{\left(R_{c}-R_{\infty}\right)}{R_0}\frac{e^{\mu r_c}}{r_c^2} \frac{MM_ce^{-\mu r}}{8\pi M_{Pl}^2r}
\end{eqnarray}
where $\frac{M_c}{8\pi M_{Pl}^2}=\frac{R_0 r_c^3}{6\beta}$.
Using the definition of a Yukawa potential (Eqn (\ref{yuk})) we find
\begin{eqnarray}\label{eq:alpha}
\alpha=12\beta^2 b\frac{\left(R_{c}-R_{\infty}\right)}{R_0}\frac{e^{r_c/\lambda}}{r_c^2}
\end{eqnarray}
and
\begin{equation}
\lambda=\frac{1}{\mu}=\sqrt{6b}.
\end{equation}

We now consider the two limiting cases of the theory. Firstly, for $\mu r_c\ll1$ ($\delta_R\rightarrow0$) we have $e^{\mu r_c}\rightarrow 1$ and 
Eqn (\ref{eqn:consts}) can be approximated by
\begin{eqnarray}
R_c&=& R_{\infty} + (\mu r_c)^2(R_{0}-R_{\infty})/3.
\end{eqnarray}
Substituting this result into our expression for $\alpha$, and assuming that $R_0\gg R_\infty$ gives
\begin{eqnarray}
\alpha=4\beta^2b\mu=\frac16.
\label{alpham2}
\end{eqnarray}
Note that while this result is independent of $b$, by assuming the conformal transformation, $f''\ne 0$ and $b\ne 0$ is implicit.

The second limiting case occurs when $\mu r_c \gg 1$ ($\delta_R\rightarrow1$) and Eqn (\ref{eqn:consts}) becomes
\begin{eqnarray}
R_c&=& \frac{R_0+R_\infty}{2},
\end{eqnarray}
and for $R_0>R_\infty$, $\alpha$ is now given by
\begin{eqnarray}
\alpha=6\beta^2 b \frac{e^{\mu r_c}}{r_c^2}.
\end{eqnarray}
In this case $\alpha$ depends on $b$ and $r_c$. For a given experimental radius, $r_c$, there is a minimum value of $\alpha$ which will occur for $b=\frac{r_c^2}{24}$.  The minimum value of $\alpha$ for a given radius is
\begin{eqnarray}
\alpha_{\rm min}=\frac{e^2}{16}=0.46.
\end{eqnarray}
Note, however, this minimum may not be seen, since $b=\frac{r_c^2}{24}$ may not occur within the regime for $b$ in which $\mu r_c\gg 1$.

It can therefore be noted that the minimum force for the complete range of $b$ will be given by $\alpha=\frac16$, when $\delta_R\rightarrow0$.   
For the upper limit of $b=10^{53}{\rm GeV}^{-2}$, $\lambda\approx 10^{11}$m or $25r_E$.  

For large $b$, such a force would be ruled out by experiment.  As $b$ is decreased, $\alpha$ increases, but since 
the range decreases, the force will eventually become undetectable. As the range is reduced, however, it becomes 
necessary to consider a smaller scale experiment. For example, laboratory experiments measuring distances less 
than $1cm$ consider a test mass in a vacuum.  This alters the experimental setup considered here and is beyond 
the scope of this paper\footnote{In addition, in reducing $b$ considerably, we would have to include the $a/R$ term in 
the theory, which makes the analysis of local constraints considerably harder.}.

\subsection{General $m$}

We continue our investigation by considering more generalised formulations of the theory, with 
\begin{equation}
f(R)=R+bR^m 
\end{equation}
where $m$ is a positive constant.  

Analytical solutions for the general case cannot be found for Eqn (\ref{fullR}).  It is therefore necessary to solve specific cases numerically.
The ``width'' of the potential minimum is described using the second derivative of the potential, as in Eqn (\ref{eqnmass}).  
As the density decreases, the mass increases and the ``width'' of the minimum decreases.
We may consider the scalar field to be moving along an inverted potential, as in bubble nucleation~\cite{Coleman:1977py} and, in this notation, the minimum should be thought of as a maximum (see \cite{Khoury:2003rn} for a good description).
Therefore, as the density decreases, the maximum increases in height but decreases in width.  
Numerically, this provides a challenge: an extreme fine-tuning of initial conditions is required in order to sit the field on such a maximum (or conversely, to place the field in its minimum).
Even with high-precision calculations, solutions are difficult to obtain.  
It is numerically easier to consider the evolution of the scalar field $\sigma$ directly and to observe the solutions as the external density is decreased.

The equivalent equation for $\sigma$ is given by
\begin{equation}\label{eq:sigmageneral}
%\frac{d^2\sigma}{dr^2}+\frac{2}{r}\frac{d\sigma}{dr}-\frac{Rf'-2f+\beta\kappa^2\rho}{3f'^2}=0.
\frac{d^2\sigma}{dr^2}+\frac{2}{r}\frac{d\sigma}{dr}-\frac13\frac{dV_{\rm eff}^{(\sigma)}}{d\sigma}=0.
\end{equation}
If we restrict ourselves to the regime where
\begin{equation}
R>mbR^m
\end{equation}
we can make the approximations
\begin{equation}
f'\sim 1 \,\,\,\,\,\,\,  \sigma\sim -mbR^{m-1}
\end{equation}
and so Eqn (\ref{eq:sigmageneral}) can be approximated by
\begin{equation}
\frac{d^2\sigma}{dr^2}+\frac{2}{r}\frac{d\sigma}{dr}+\frac{1}{3}\left(\frac{-\sigma}{mb} \right)^{\frac{1}{m-1}}-\frac{\beta\kappa^2\rho}{3}=0.
\label{sigmafull}
\end{equation}
Note that $\sigma$ is negative and hence the third term is real.
In general it is not possible to find analytic solutions to this equation, but numerical solutions are possible.  
As for $R(r)$, we shall denote the minima of $\sigma$ internally and externally as $\sigma_0$ and $\sigma_\infty$ respectively.  At $r=0$, $\sigma=\sigma_i$.

\subsubsection{Fifth Force}
Heuristically, with knowledge from the $m=2$ case, it is possible to estimate how the strength of the fifth force will change with $b$.  When $\delta_\sigma\rightarrow0$, we might expect the strength to take a value independent of $b$.  As $b$ is decreased, $\delta_\sigma$ increases until $\delta_\sigma=1$.  As this happens, the difference between minima ($\sigma_0-\sigma_\infty$) becomes larger, while the transition length becomes shorter.  This implies that the resultant force becomes stronger, but the interaction range decreases.

From this argument, the most detectable force will occur for the maximum allowed value of $b$, when $\delta_\sigma\rightarrow0$.  We will initially consider this case, in order to calculate the strength and range of the force.

We will consider four distinct regimes of Eqn (\ref{sigmafull}); one internal solution ($r<r_c$), and three external solutions ($r>r_c$), which we shall denote by I,II,III and IV respectively. These correspond to neglecting combinations of terms in Eqn (\ref{sigmafull}).

Let us consider the internal solution first. If $\delta_\sigma\rightarrow0$, the system is heavily damped and Eqn (\ref{sigmafull}) can be approximated by
\begin{equation}
\frac{d^2\sigma}{dr^2}+\frac{2}{r}\frac{d\sigma}{dr}-\frac{\beta\kappa^2\rho}{3}\approx 0.
\end{equation}
A solution can be found
\begin{equation}
\sigma_{I}(r)=\frac{\beta\kappa^2\rho r^2}{18}+\sigma_i
\end{equation}
where we have used the condition that $\frac{d\sigma}{dr}=0$ when $r=0$.
Note once again that $\sigma$ is negative and hence the magnitude of $\sigma$ decreases.
Outside $r=r_c$, the density drops and the system can initially be re-expressed by
\begin{equation}
\frac{d^2\sigma}{dr^2}+\frac{2}{r}\frac{d\sigma}{dr}\approx 0.
\end{equation}
which has a trivial solution
\begin{equation}
\sigma_{II}(r)=\frac{-A}{r}+C.
\label{sigmaII}
\end{equation}
Matching the derivatives at $r=r_c$ of $\sigma_I$ and $\sigma_{II}$ leads to an expression for $A$:
\begin{eqnarray}
A&=&\frac{\beta\kappa^2\rho_c r_c^3}{9}
\end{eqnarray}
Assuming that $\lambda$ is small and using Eqns (\ref{yuk}) and (\ref{V_E}), $A$ can be related to $\alpha$:
\begin{eqnarray}
\alpha=\frac{6\beta A}{\kappa^2\rho_c r_c^3} = \frac{2}{3}\beta^2 = \frac{1}{6}.
\label{alphageneral}
\end{eqnarray}
Hence, in this limit, the strength of the fifth force is seen to be independent of $m$ or $b$.
This value for $\alpha$ is the same limit as seen in Eqn (\ref{alpham2}) for $m=2$. 

The third and fourth regimes occur when the potential terms (last two terms on the LHS of Eqn (\ref{sigmafull})) 
become important. Firstly, consider that both terms are of similar magnitude (regime III).  We can approximate 
the radius of decay from $\alpha=\frac16$ by the interaction range of $\sigma$, $\lambda=1/\mu$, where $\mu$ is given 
by Eqn (\ref{eqnmass}). As the density decreases, the mass increases for $m<2$ and decreases for $m>2$.  The 
interaction range, $\lambda$, does the converse.

Finally, consider the case in which the density is low enough that the final term can be ignored:
\begin{equation}
\frac{d^2\sigma}{dr^2}+\frac{2}{r}\frac{d\sigma}{dr}+\frac{1}{3}\left(\frac{-\sigma}{mb} \right)^{\frac{1}{m-1}}=0.
\end{equation}
Then the solution starts to decay when then the second and third terms are of the same magnitude.
We make the assumption that $\sigma(r)$ follows $\sigma_{II}$ up until this time, where $C$ is related to the density and is negligible.
It is a simple rearrangement to show that the two terms roughly equate when 
\begin{equation}
\left(\frac{r}{r_c}\right)^{3m-4}\approx\frac{6mb}{r_c^2}\left(\frac{2}{3}\beta\kappa^2\rho_c\right)^{(m-2)}
\label{minradius}
\end{equation}
This radius sets the length of decay.

Note that regimes III and IV are exclusive: either the density term contributes or not. 
If the density term contributes, the solution follows regime III and the decay scale is set by the mass term.
If the density is below a threshold, regime IV, it can no longer affect the decay length, which is then set by Eqn (\ref{minradius}).
Hence this latter radius sets the minimum decay length and we shall call it $r_{\rm min}$.
Between $r_c$ and $r_{\rm min}$, we expect the strength of the force to be $\frac16$.
Table \ref{tab2} gives the minimum radius of decay for varying $m$ in the Earth system when $b$ is close to the upper limit.

\begin{table}
\begin{center}
\begin{tabular}{ccc}
\hline
$m$&$b$&$r/r_E$ \\ \hline
$\frac32$&$10^{25}$&$10^{15}$ \\
$2$&$10^{52}$&$10^4$ \\
$\frac52$&$10^{78}$&$10^2$ \\
$3$&$10^{105}$&$10^1$ \\
\end{tabular}
\caption{Minimum decay radius for the Earth with varying $m$ given by Eqn (\ref{minradius}).  Values for $b$ are taken close to the upper limit and correspond to the numeric parameters in Figure (\ref{forcediags}).}
\label{tab2}
\end{center}
\end{table}

In order to verify this behaviour, we ran numerical simulations within the Einstein frame, using Eqn (\ref{sigmafull}).  We ran the simulation for varying $m$, once again using values for $b$ close to the maximum range.
The code explicitly solved for $\sigma(r)$ and $d\sigma(r)/dr$ allowing for direct calculation of the field's force, $F_{\sigma}$. 

As in Section \ref{secm=2}, we chose the Earth system. For each choice of $m$, the 
external density was varied to observe the effect on the decay length.  
A physical situation is one in which the external density is that of the solar system, 
$\kappa^2\rho_{SS}\approx 10^{-78} {\rm GeV}^2$.  

\begin{figure}[ht]
\begin{center}
\scalebox{0.5}{\includegraphics{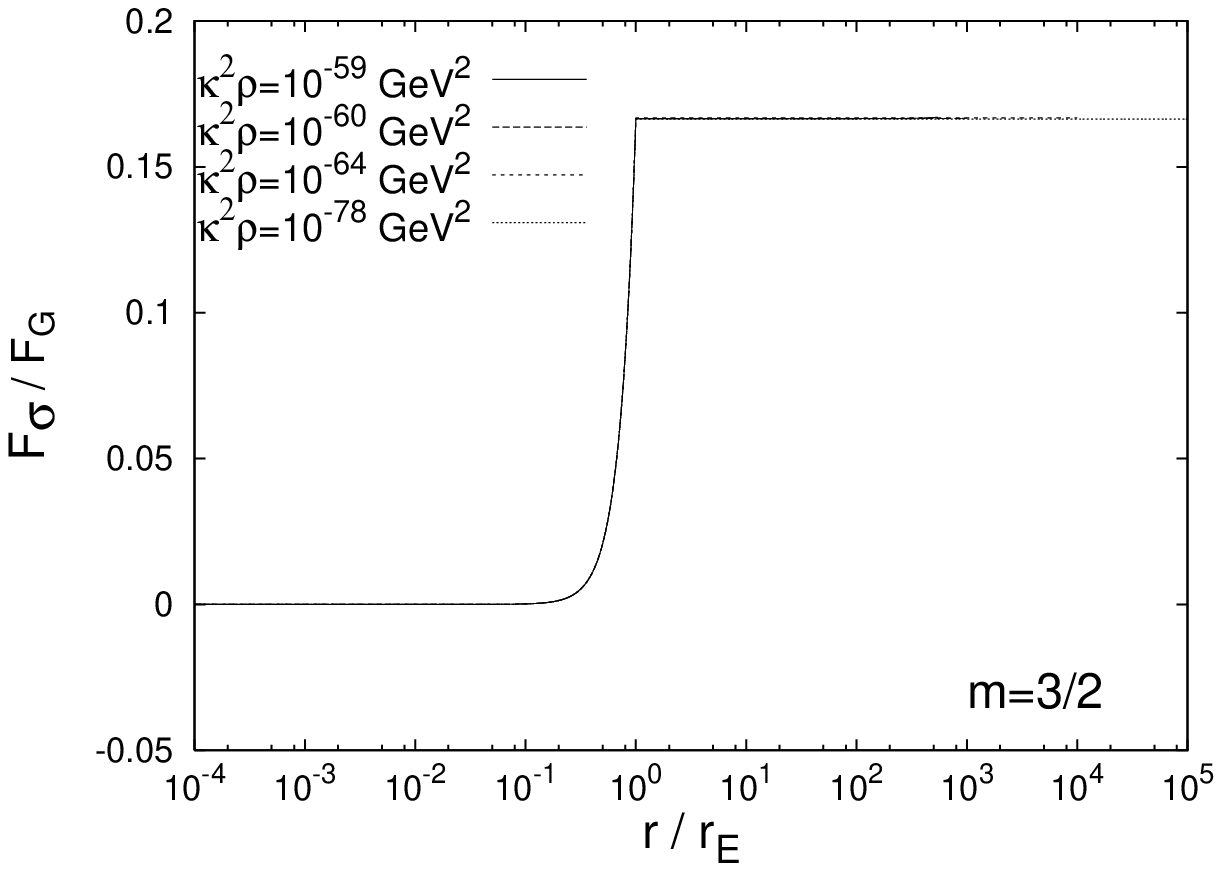}}
\scalebox{0.5}{\includegraphics{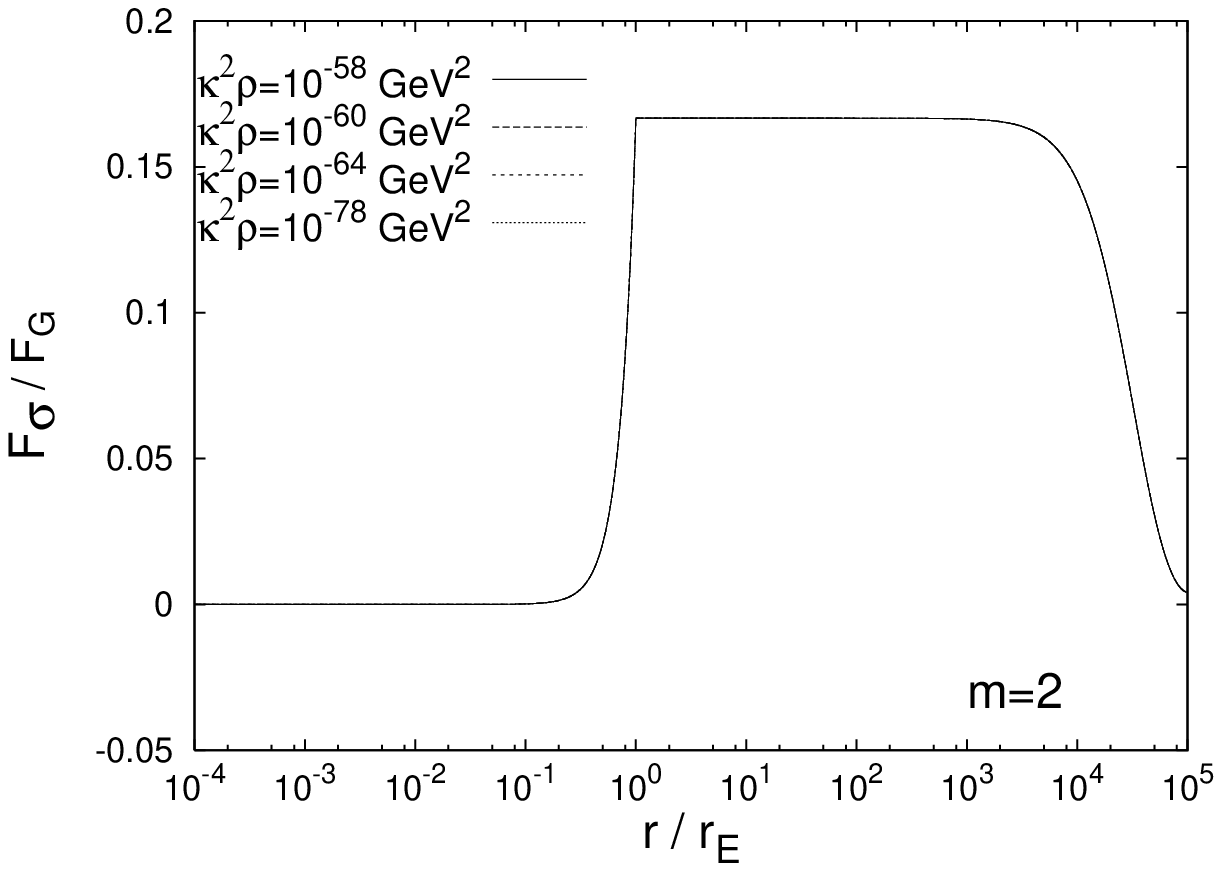}}
\scalebox{0.5}{\includegraphics{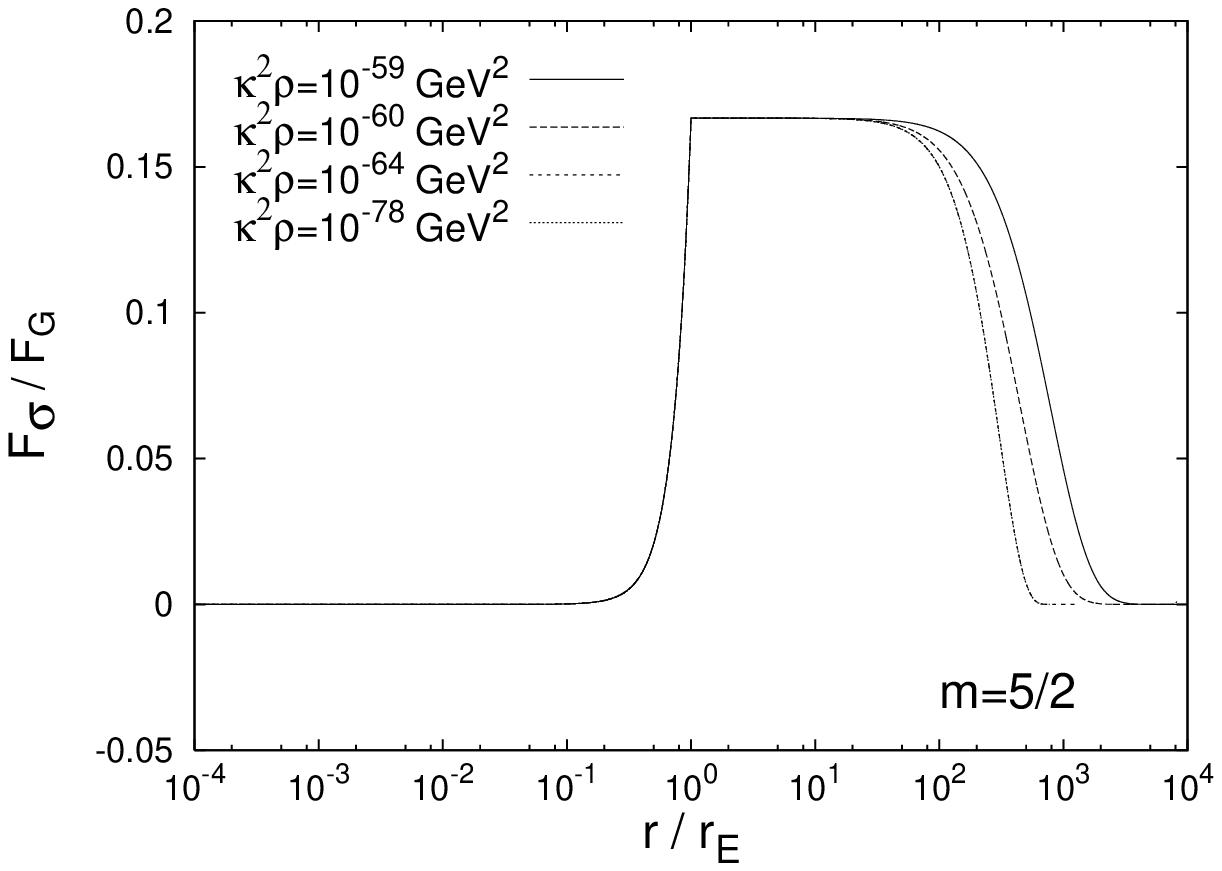}}
\scalebox{0.5}{\includegraphics{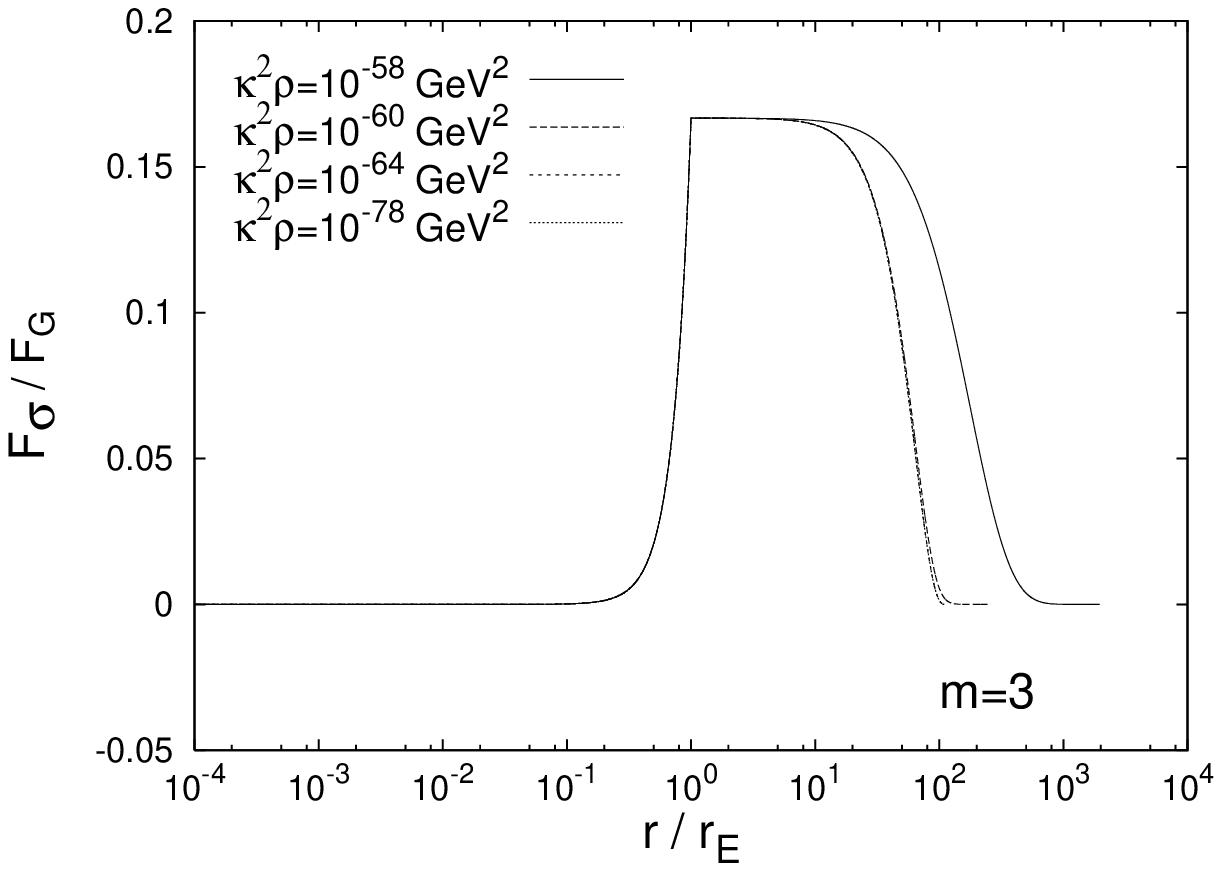}}
\caption{\label{forcediags}The ratio of forces due to the scalar field, $\sigma$ and Newtonian potential, shown for different $m$.}
\end{center}
\end{figure}

The results are shown graphically in Figure (\ref{forcediags}).  
Numerically, a binary search was undertaken in order to find the $\sigma_i$ value which meant the external solution reached $\sigma_\infty$.  
No assumption was made as to the value of $\sigma_i$.  
However, in all cases, $\sigma_i\ll \sigma_0$, leading to models in the thickshelled regime (as expected for large $b$ values).  
The analytics described above should be a good approximation to the numerical solution.

Regimes I and II given analytically above are easily seen for all choices of $m$.  
From Table \ref{tab2}, it is justified that we do not see a decay for $m=3/2$ for the range of $r$ in our simulation.  Only regimes I and II are visible and the force remains constant outside $r=r_c$.
For $m=2$, the decay radius occurs around $r\approx 8000$, which is approximately equal to $\lambda=1/\mu$ from Section \ref{forcem=2}.  Note that the decay radius is independent of the density, as expected from the mass term given in Section \ref{jordanm=2}.
Regimes III and IV are explicitly seen for $m=5/2$ and $m=3$.  As the density decreases, the decay length also decreases, until the density finally has no more effect.
The minimum decay length matches Eqn (\ref{minradius}) within an order of magnitude.

Previously we argued heuristically that the force is a maximum when $b$ takes its maximum value.  In order to test this, numerical solutions were found as $b$ decreased away from this regime.  
Due to numerical difficulty, it was not possible to simulate the complete range of $b$ for each choice of model $m$.  Instead $b$ was increased over an order of magnitude to test the effect on $\alpha$.

\begin{figure}[t!]
\begin{center}
\scalebox{0.5}{\includegraphics{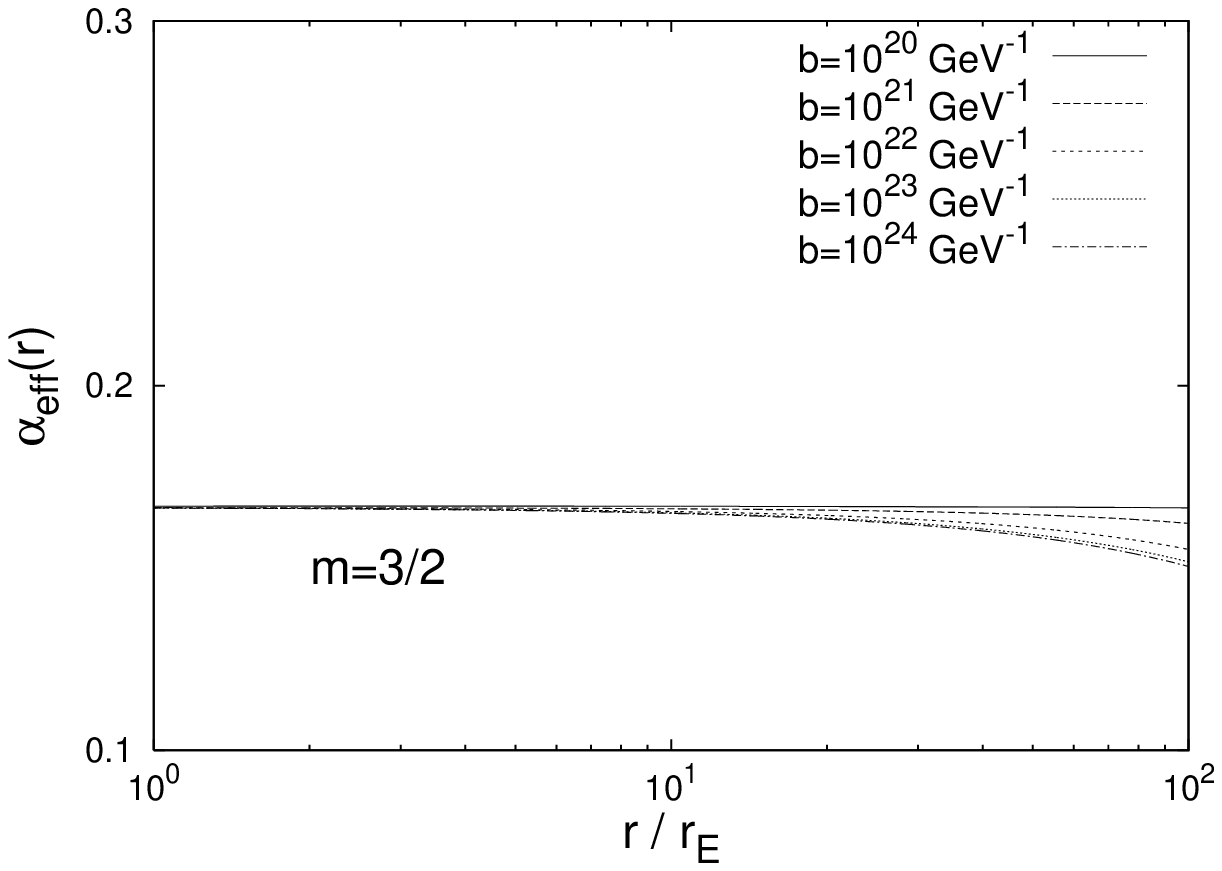}}
\scalebox{0.5}{\includegraphics{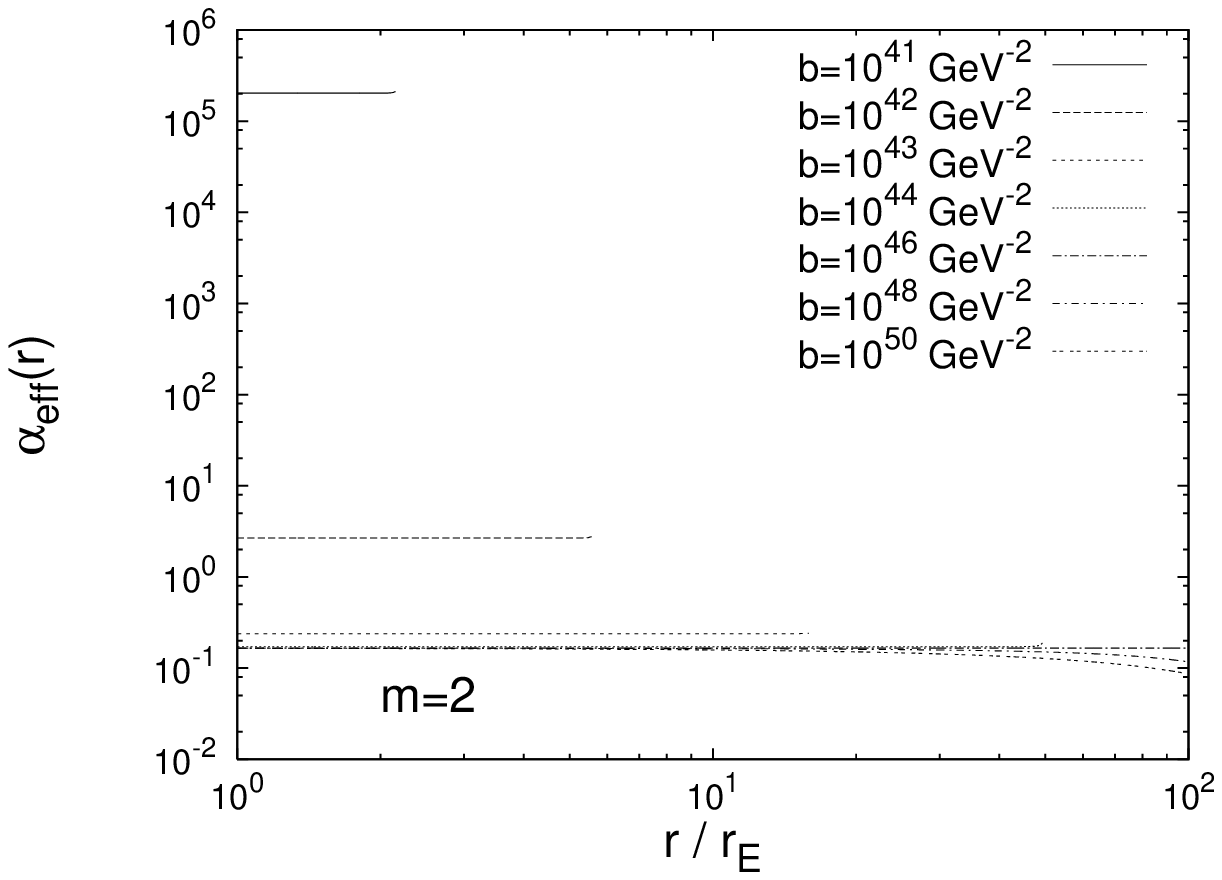}}
\scalebox{0.5}{\includegraphics{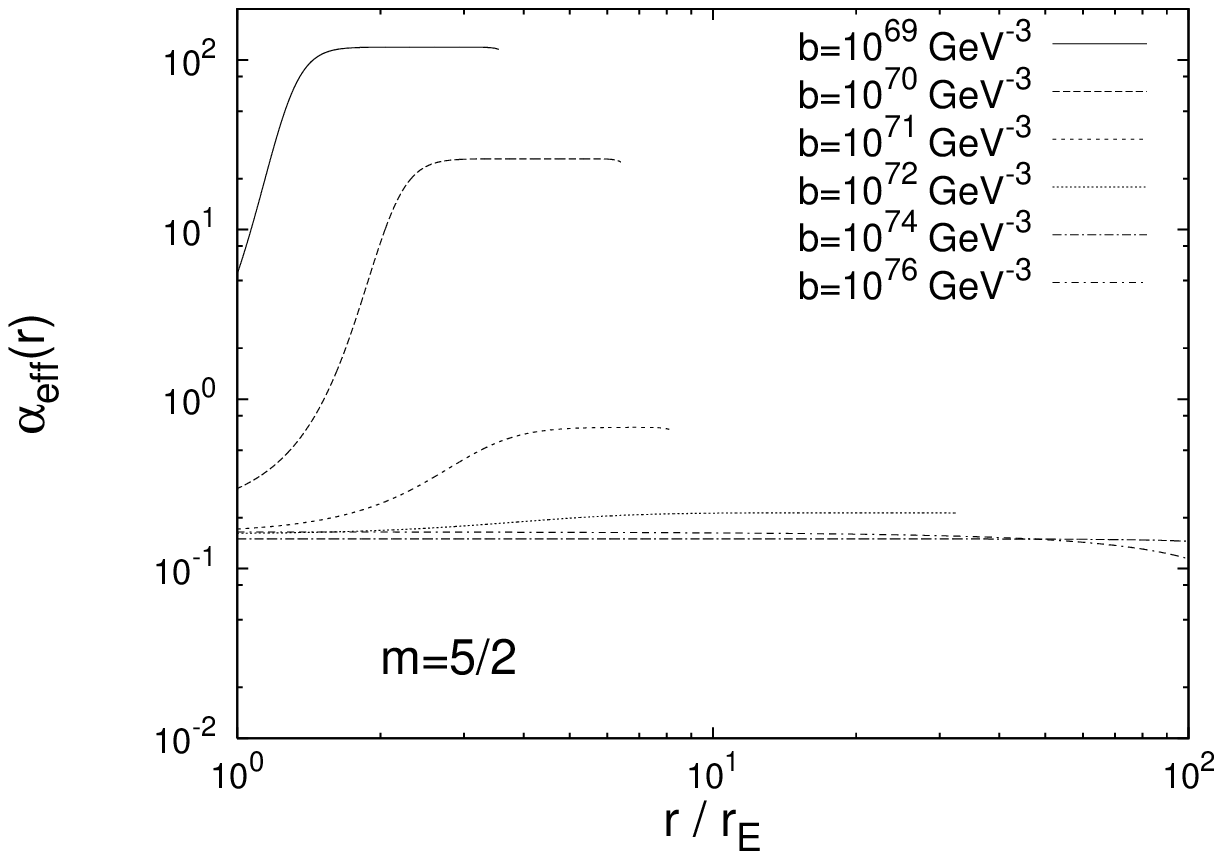}}
\scalebox{0.5}{\includegraphics{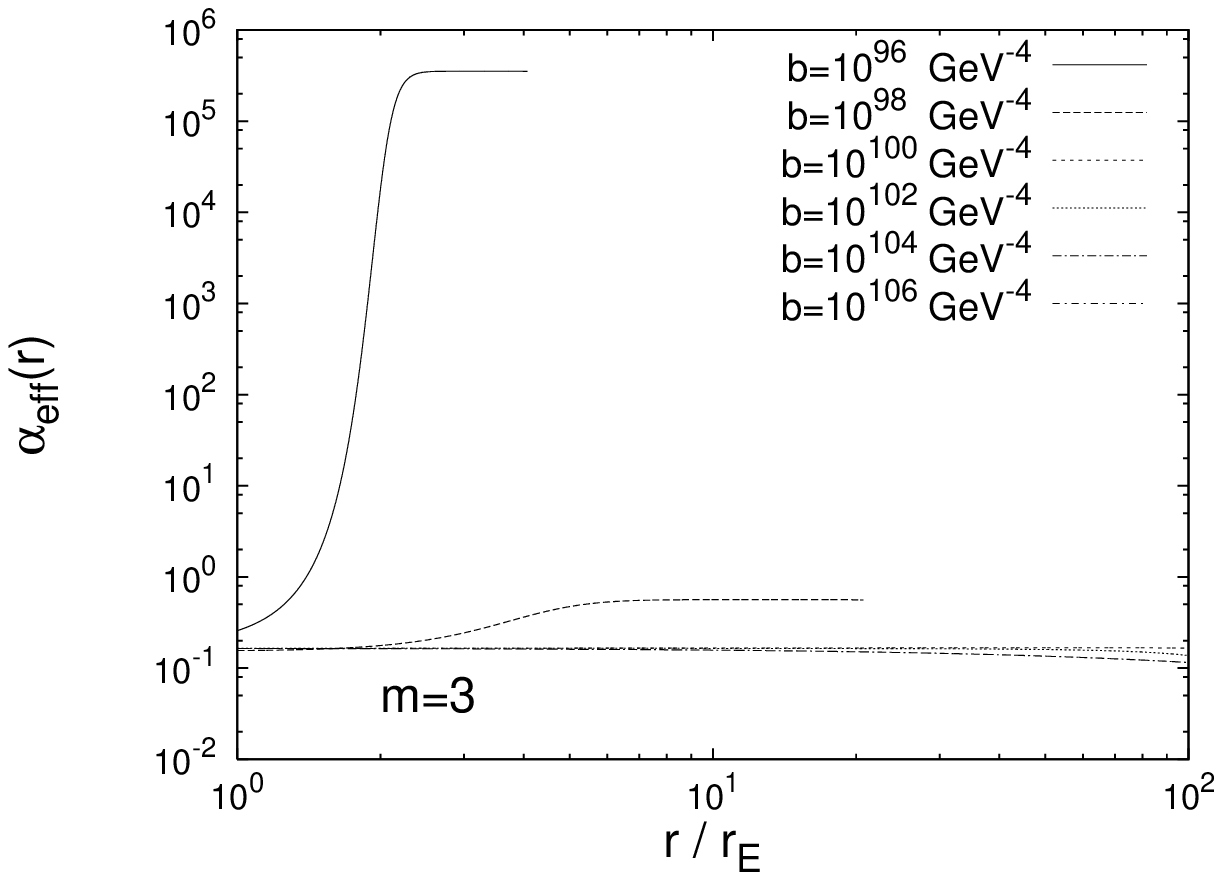}}
\caption{\label{alphadiags}Numerical calculation of the effective field strength, $\alpha_{\rm eff}$ as a function of radius.  A constant $\alpha(r)$ denotes a pure Yukawa regime.  $R_E=10^{-53}{\rm GeV}^2$,  $R_\infty=10^{-55}{\rm GeV}^2$.
For the numerics shown: $\lambda/r_E=24-7600$($m=\frac32$), $\lambda/r_E=10^{-2}-700$($m=2$), $\lambda/r_E=10^{-2}-70$($m=\frac52$), $\lambda/r_E=10^{-1}-3800$($m=3$).}
\end{center}
\end{figure}

In the general $m$ model, it is not possible to show analytically that the force has Yukawa form. 
Indeed, the evolution of $\sigma$ does not fit perfectly to a Yukawa field, but an approximation can be made. 
We will describe the force given by an effective strength, $\alpha_{\rm eff}$, which varies with radius.  We define
\begin{equation}
\sigma(r)=\frac{-\alpha_{\rm eff}(r)~m_c~e^{-\mu(r)r}}{\beta~8\pi M_{Pl}^2~r}+\sigma_0.
\label{effsigma}
\end{equation} 
where $\sigma(r)$ is calculated numerically.
A pure Yukawa field will yield a constant $\alpha_{\rm eff}$.
The numerics directly calculate the first and second field derivatives with respect to $r$, $\sigma^{\prime}$ and $\sigma^{\prime\prime}$ and the effective mass, $\mu(r)$, is given by
\begin{equation}
\mu(r)=-\frac{1}{r}-\frac12\frac{\sigma^{\prime\prime}}{\sigma^{\prime}}
    \pm \frac12\frac{\sqrt{-4 + \left(\frac{\sigma^{\prime\prime}}{\sigma^{\prime}}r\right)^2}}{r}
%\left(\right\frac{\sigma^{\prime\prime}}{\sigma^{\prime}})^2}}{r}
\end{equation}
where we take the positive root.  It is straightforward to rearrange Eqn (\ref{effsigma}) to determine $\alpha_{\rm eff}(r)$ numerically.  
The results are shown graphically in Figure (\ref{alphadiags}) for $m=\frac32,2,\frac52,3$.

When $b$ takes a maximum value (allowed by the condition for a minimum), $\delta_\sigma\rightarrow0$ and $\alpha(r)=\frac16$.  The tails at high $r$ values correspond to the field evolving from the minimum due to extreme fine-tuning of initial conditions.  
At this point, the field no longer follows a Yukawa potential.
As $b$ is decreased from its maximum, the effective field strength increases, as expected.  
For $m=2$, $\alpha_{\rm eff}$ is given to high precision by Eqn (\ref{eq:alpha}).
For cases approaching the thinshell regime, the field decays quickly, reaching $\sigma_\infty$ in shorter and shorter lengths.  
The numerical calculation of $\mu(r)$ agrees very accurately with the mass given in Eqn (\ref{eqnmass}).  
Since the interaction range is the inverse of $\mu$, it is directly proportional to $\sqrt{b}$ and shortens with decreasing $b$.

The ranges of $\lambda$ for the numerical results are given with Figure (\ref{alphadiags}). The minimum interaction range shown is quoted as $10^{-2}r_E$, which corresponds to $69km$.
As $b$ decreases, it may become possible to hide the field, since even though the field strength increases, the interaction range decreases.

Note that until this point we have neglected the Earth's atmosphere.  
If the field always stays close to $R_{SS}$ (thick regime), the presence of the atmosphere makes little effect on the field and we may neglect the atmosphere in this regime.  We have checked this assumption numerically in the case $m=2$.
If $R_i\approx R_0$ (thin regime), the field rolls quickly from its minimum within the Earth to that of the Solar System.  The transition length can be the same magnitude of the atmospheric depth and we would expect the field to sit at
its minimum, both within the Earth and the atmosphere.  
The fifth force
constraints upon this system would therefore arise from both the
Earth-atmosphere boundary and the atmosphere-space boundary.  Given
that we have shown that $\alpha\geq 1/6$, independent of the values of
$R_{0}$ and $R_\infty$, we can assume that
our results will still hold.  

Once again, as in the $m=2$ case, when $\lambda$ becomes very small, local experiments testing a fifth force require a simulation of test cases in vacua, which is outside the scope of this paper.

\section{Dark Energy and $f(R)=R-a/R^n+bR^m$}
\label{secDE}
As mentioned already earlier, it has been suggested that a modified gravity theory with both positive and negative 
curvature terms can provide a mechanism for early and late time inflation, while hiding a local fifth force arising 
from the additional scalar degree of freedom. In this section, we consider the effect of these terms, both locally 
and cosmologically, and investigate the allowed parameter range of $b$. 

\subsection{Early Time Inflation}
At early times in the universe, with extreme high curvature, the theory tends to the limit $R+bR^m$.  
It has been postulated that the higher curvature terms may drive early time inflation (c.f. $R^2$ 
inflation~\cite{Starobinsky:1980te}). 
It is possible to estimate the required value for $b$, such that fluctuations in the 
Cosmic Microwave Background assume the observed amplitude and spectral index.  For 
$m=2$, $b\approx 10^{-24} {\rm GeV}^{-2}$~\cite{Hwang:2001pu}. Of course, it should be remembered that, 
whilst it is advantageous that this theory may lead to early and late time inflation, it is certainly 
not a requirement.  Therefore $b$ need not be constrained by such results.  We merely mention it for 
completeness.

\subsection{Locally}
In regions of high curvature, $f(R)$ can be approximated by $R+bR^m$ as in Section \ref{seclocal}.  As 
the curvature decreases, the $a$ term becomes more significant, which eventually results in the loss of 
exterior minima.  We have confirmed numerically that, provided an exterior minimum exists, the evolution 
of $R(r)$ follows exactly the results obtained in Section \ref{seclocal}, even when $a\ne 0$.

In Section \ref{sectheory}, it was stated that $f''>0$ for stability and locally
\begin{eqnarray}
\frac{a}{bR^{(n+m)}} = \frac{(10^{-42})^{~2(n+1)}}{b~R^{~(n+m)}}~{\rm GeV}^{~2(1-m)} < 1.
\label{ignorea}
\end{eqnarray}
For the Earth, the field assumes a minimum $R_E\approx 10^{-53}{\rm GeV}^2$ and the existence of minimum
\footnote{For $m<2$, the upper limit is not imposed due to existence of minima, but we will see later that Big Bang Nucleosynthesis constraints force this limit.}
 leads to an upper limit on $b$ (Eqn (\ref{bupperlim})). Hence
\begin{eqnarray}\label{brangeE}
10^{\left(53m-31n-84\right)}~{\rm GeV}^{2(1-m)}~<~b~<~(10^{-53})^{~(1-m)}~{\rm GeV}^{2(1-m)}
\end{eqnarray}
within orders of magnitude.
In the solar system, assuming $R$ sits in the minimum, $R_{SS}\approx 10^{-78}{\rm GeV}^2$ and therefore
\begin{eqnarray}\label{brangeSS}
10^{\left(78m-6n-84\right)}~{\rm GeV}^{2(1-m)}~<~b~<~(10^{-78})^{~(1-m)}~{\rm GeV}^{2(1-m)}
\end{eqnarray}
It is straightforward to show that, for the cases of $m$ and $n$ we consider, there is no overlap between the upper limit of 
Eqn (\ref{brangeE}) and the lower limit of Eqn (\ref{brangeSS}). Therefore, for a given $b$, we cannot simultaneously find 
minima within the Solar System and the Earth.

In the absence of an external minimum, we are unable to specify the value that the field takes at infinity.  In this case, we can either assume an external field value, or say nothing regarding the evolution of $R(r)$.

We shall consider two exclusive regimes; we take $b$ such that (a) a minimum exists in the solar system and (b) a minimum exists in the Earth. 

\begin{itemize}

\item[(a)] For a minimum to exist in the solar system, the value of $b$ is high.  
For this range of $b$, the inverse curvature plays no significant role and the results of Section \ref{seclocal} hold.  Due to the large value of $b$, the theory is restricted to a thickshell regime, which implies $R\sim R_\infty$ always.  In this case, we expect $\alpha=\frac16$ within a minimum radius given by Table (\ref{tab2}).

\item[(b)] We consider a value of $b$ so that a minimum exists on Earth and locally the theory takes the limit $R+bR^m$.  
For a given $b$ value, there will be an exterior field cut-off, $R_{\Lambda}$ for the existence of minima.  For $R$ above this cut-off, the theory is well described by the theory discussed in Section \ref{seclocal}.  Again, we can expect $\alpha>\frac16$.

Below the field cut-off, no exterior minimum exists and we are unable to predict $R_\infty$.  If we assume, however, that the field sits at a value 
$R_{\rm ext}$ (N.B. not given by the minimum) , we can make some predictions about the theory.
Firstly, if $R_{\rm ext}>R_\Lambda$ then we have the same case as in the paragraph above.  Secondly, if $R_{\rm ext}<R_\Lambda$, the theory tends to $R-a/R^n$ at large distances from the Earth and this is ruled out experimentally~\cite{Chiba:2003ir}.

\end{itemize}

\noindent For a $b$ value outside of these two regimes, we are unable to say where the field will sit at any given radius and we cannot calculate the strength of the force.

\subsection{BBN Constraints on $b$}
\label{secBBN}
One of the basic pillars of modern cosmology is Big Bang Nucleosynthesis (BBN) and the modified gravity theory we consider has clearly to 
reproduce the observed light element abundances. 

Stringent limits on the abundances of $^3$He and 
$^7$Li lead to a constraint on the deviation of the Hubble parameter during BBN from standard cosmology. 
Generally, a non-standard expansion rate can be parameterised by an expansion rate factor $S=\frac{H}{H_{GR}}$, 
where $H$ is the expansion factor in the physical (Jordan) frame and $H_{GR}$ is the expansion rate in Einstein's 
theory for the same matter content. This in turn can be related to the number of extra relativistic neutrino species, 
$\Delta N_{\nu}$ \cite{Kneller:2004jz}, by
\begin{eqnarray}
S=\left(1+\frac{7\Delta N_{\nu}}{43} \right)^{1/2}
\label{S}
\end{eqnarray}
It then follows that, for a deviation from General Relativity
\begin{eqnarray*}
\frac{H^2-H_{GR}^2}{H^2}= 1-\frac{1}{S^2}=\frac{7\Delta N_{\nu}/4}{10.75+7\Delta N_{\nu}/4}.
\end{eqnarray*}
In the Jordan frame, the variation of the Hubble parameter can be related to the 
conformal factor $A=\exp(\sigma/2)$~\cite{Bartolo:1999sq}
\begin{eqnarray}\label{basicbbn}
\frac{\Delta H_{JF}^2}{H^2_{JF}}=1-\frac{A^2_0}{A^2_{\rm BBN}} 
\end{eqnarray}
Using Eqn (\ref{sigmadef}) we find
\begin{eqnarray}
\label{limits}
0.86 < \frac{f^{\prime}(R_{\rm BBN})}{f^{\prime}(R_0)} < 1.19
%1.0 \geq&  \frac{1+a\alpha R^{-(\alpha+1)}+ b\beta R_{\rm BBN}^{\beta+1}}{1+a\alpha R^{-(\alpha+1)}+ b\beta R_{\rm BBN}^{\beta+1}} &\geq 0.86
\end{eqnarray}
where we have used $|\Delta N_{\nu}| < 1$ \cite{Kneller:2004jz,Copi:1996pi}. The method above has been used 
to look for deviations of BBN in scalar--tensor theories. 

During the BBN epoch, we expect positive curvature terms to dominate, due to the high curvature\footnote{Deviating too much from 
this might result in a non--standard cosmological evolution and is ruled out \cite{Amendola:2006kh}.}.
At the present time, however, the observation of dark energy leads us to assume that the term involving 
inverse powers of $R$ is beginning to dominate.  These assumptions lead to the following constraint:
\begin{eqnarray}
\frac{1+bmR_{\rm BBN}^{m-1}}{1+anR_0^{-(n+1)}} \sim 1,
\end{eqnarray}
where we have assumed that during BBN the $a/R$ term is negligible and that the $bR^m$ term is negligible for 
the cosmological evolution at the present epoch. 
If we require that acceleration today arises due to the inverse curvature term, the terms in the denominator must be of order unity.  Hence, the $b$ term must be order unity for the BBN constraint to hold.  This leads to a fine-tuning of the parameter $b$:
\begin{eqnarray}
b \sim R_{\rm BBN}^{1-m}.
\end{eqnarray}
Note that this is a product of the coincidence problem of dark energy; the fact that the $a$ term just starts to dominate today implies that the $R^m$ term stops dominating at the BBN epoch.
Since, we expect $R$ to decrease since the BBN epoch, we can assume that such positive curvature terms will be sub-dominant after this time ($bR^{m-1}<1$).  

Therefore the evolution of the universe must closely resembles that produced by Einstein gravity at times from BBN until today. 
We can therefore approximate $R= \beta\kappa^2\rho$, which is the condition for the field to be sat close to its minimum 
(see Eqn (\ref{minexist})). With this argument, we can estimate $R_{\rm BBN}\sim \kappa^2\rho_{0}(1+z_{\rm BBN})^3$.  Hence,
\begin{eqnarray}
b \sim (10^{-54}~{\rm GeV}^2)^{1-m}.
\label{BBNb}
\end{eqnarray}
Note that this value lies just inside the range for the existence of a minimum on the Earth.

\subsection{Present Day Acceleration}

As argued in the previous section, we should assume for cosmological evolution that the field sits close to its minimum (and $f(R)\approx R$) until today when inverse curvature terms begin to play a role.
It is interesting to consider the constraint on $b$ coming from this requirement.  Note that, at recent times, we expect the field to move out of its minimum in order to induce acceleration.

We numerically solve Eqn (\ref{effectivepotential}) in order to find $R$, using the full $f(R)=R-a/R^n+bR^m$ theory.  We always take the positive, non-zero root, since $R>0$.
Using this value, we solve Eqn (\ref{stability}) to find the limits on $b$.  
The current cosmological
energy density is $\rho_{0} = 1.3\times 10^{-47}${\rm GeV}$^4$ and we assume $a$ to take the value in Eqn (\ref{avalue}).
The limits for the cases we have considered are:
\begin{eqnarray}
m=&3/2 & ~~~b>10^{42}{\rm GeV}^{-1} \\
m=&2   & ~~~b>10^{83}{\rm GeV}^{-2} \\
m=&5/2 & ~~~b\sim 10^{125} {\rm GeV}^{-3} \\
m=&3   & ~~~b\sim 10^{166} {\rm GeV}^{-4}.
\end{eqnarray}
Note that for $m\leq2$, the upper limit in Eqn (\ref{stability}) is always satisfied, while for $m>2$, the 
limits tightly constrain $b$. We also find that these limits are largely independent of
$n$, over the range $n=\frac{1}{2},...,2$.

\section{Comparison of Regimes}
\label{seccomp}

In this paper, we have looked at the many various constraints that can be placed on a modified gravity theory, $f(R)=R-a/R^n+bR^m$.  These constraints include local fifth force experiments, Big Bang Nucleosynthesis and cosmological evolution.  In addition, we compare to early time inflationary results.

For densities corresponding to the Earth, solar system and cosmology ($\rho_E$,$\rho_{SS}$,$\rho_0$), there is no value of $b$ that simultaneously satisfies the existence of a minimum in all media.  Indeed, $b$ values exist for which there is an absence of minima in all these regimes.  For these values, it is not possible to comment on the evolution of $R$ either locally or cosmologically.

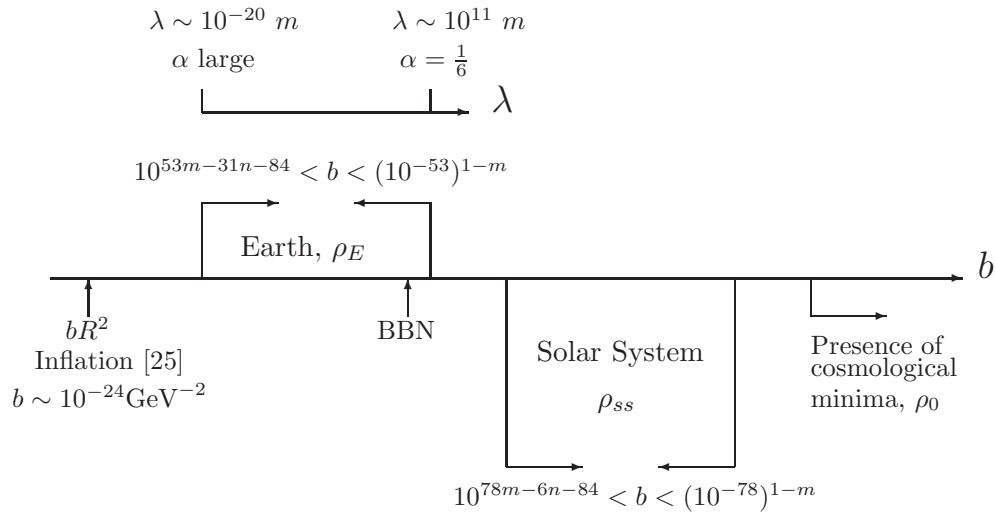
\begin{figure}[h]
\setlength{\unitlength}{1cm}
\begin{picture}(13.5,7)
%b line
   \put(1,3){\vector(1,0){12}}
   \put(13.2,3.0){\large{$b$}}
%inflation
   \put(1.5,2.5){\vector(0,1){0.5}}
   \put(1.2,2.2){\footnotesize{$bR^2$}}
   \put(0.8,1.8){\footnotesize{Inflation \cite{Hwang:2001pu}}}
   \put(0.5,1.3){\footnotesize{$b\sim 10^{-24}{\rm GeV}^{-2}$}}
%earth constraint
   \put(3.0,4.0){\line(0,-1){1}}
   \put(3.0,4){\vector(1,0){1}}
   \put(2.1,4.3){\footnotesize{$10^{53m-31n-84}<b<(10^{-53})^{1-m}$}}
   \put(3.5,3.3){{Earth, $\rho_E$}}
   \put(6.0,4){\line(0,-1){1}}
   \put(6.0,4){\vector(-1,0){1}}
%solar system constraint
   \put(7.0,0.5){\line(0,1){2.5}}
   \put(7.0,0.5){\vector(1,0){1}}
   \put(6.3,0.0){\footnotesize{$10^{78m-6n-84}<b<(10^{-78})^{1-m}$}}
   \put(7.4,1.9){{Solar System}}
   \put(8.2,1.3){{$\rho_{ss}$}}
   \put(10.0,0.5){\line(0,1){2.5}}
   \put(10.0,0.5){\vector(-1,0){1}}
%cosmol constraint
   \put(11,2.5){\line(0,1){0.5}}
   \put(11,2.5){\vector(1,0){1}}
   \put(11,2.0){\footnotesize{Presence of}}
   \put(11,1.7){\footnotesize{cosmological}}
   \put(11,1.3){\footnotesize{minima, $\rho_0$}}
%lambda line
   \put(3.0,5.2){\vector(1,0){3.5}}
   \put(6.8,5.2){\large{$\lambda$}}
   \put(2.6,5.8){\footnotesize{$\alpha$ large}}
   \put(3.0,5.2){\line(0,1){0.3}}
   \put(2.3,6.3){\footnotesize{$\lambda\sim 10^{-20}~m$}}
   \put(5.6,5.8){\footnotesize{$\alpha=\frac16$}}
   \put(5.5,6.3){\footnotesize{$\lambda\sim 10^{11}~m$}}
   \put(6.0,5.2){\line(0,1){0.3}}
%BBN
   \put(5.7,2.5){\vector(0,1){0.5}}
   \put(5.3,2.2){\footnotesize{BBN}}
\end{picture}
\caption{Schematic diagram showing the relations between regimes for $b$,$\lambda$ and $\alpha$.}
\label{diagram}
\end{figure}

We may choose $b$ such that a minimum exists for one or other of these densities.  These cases were discussed individually in Section \ref{secDE} and summarised schematically in Figure (\ref{diagram}).  
We now discuss these regimes with reference to each other to judge the consistency of the model.

Firstly, we can consider large values of $b$ such that a cosmological minimum exists until very recently.  At high curvature, for example on Earth, such large $b$ values result in a $bR^m$ theory. This theory has been considered previously~\cite{Capozziello:2004km}, where it was concluded that experimental observations require $0.25<m<1$.  

Secondly, when a minimum exists in the solar system, the regime is one of a thickshell, which can be ruled out for the models considered due to $\alpha=\frac16$ and the minimum decay radius in Table \ref{tab2}.

Thirdly, the complex consideration of minima on Earth was discussed in Section \ref{secDE}.  
The most viable values for $b$ exist at the lower end of the Earth bounds, when $\lambda$ is small enough to hide the scalar degree of freedom.  
However, it
should be noted that the arguments presented in this paper rely upon a
specific (large scale) experimental setup.  To constrain the theory on
scales $\lambda<1 \mathrm{m}$ would require a small scale experimental setup to
be considered, (i.e. a system of test-mass, vacuum flask, and
exterior medium).  
Furthermore, we can compare these lower $b$ values with the BBN result in Eqn (\ref{BBNb}).  The two constraints are inconsistent.  

It is of course possible that a viable cosmology could be produced
without assuming that the field sits in the minimum of its effective
potential. 
However numerically solving such a system is an extremely
difficult problem, and we have been unable to produce a realistic cosmology.  
For a specific model ($m=2$,$n=1$), it was noted that if the field starts with large $R$ (at early times), the 
system was not found to evolve into its local minimum~\cite{Amendola:2006kh}, see however \cite{Capozziello:2006dj}.
Note that the BBN result in Section \ref{secBBN} specifies that the field must exist close to its minimum from 
that epoch until recently.  This is inconsistent with a cosmology in which the field does not settle to its minimum.

Further to this discussion, if the model is required to produce early time inflation, none of the regimes above 
are consistent with the result found in \cite{Hwang:2001pu}.

\section{Conclusions}
\label{secconc}
In this paper, we have investigated the parameter range of the modified theory of gravity, $f(R)=R-a/R^n+bR^m$.  

For the specific case in which $a=0$, we modelled fifth force experiments using the local Earth system 
and calculated the strength and range of such a force. We find that, assuming a Yukawa form for the potential, 
the force has $\alpha>\frac16$ and $\lambda=\frac1\mu$ where $\mu$ is approximated by Eqn (\ref{eqnmass}).  This 
restricts the allowed parameter space for $b$ such that the interaction range is smaller than experimentally allowed ranges.

For the general model ($a\neq0$), we assume that $a/R^n$ leads to late time acceleration, setting $a$ for this 
to be the case.  We considered the allowed range of $b$ required by the following regimes: early and late-time 
acceleration, Big Bang Nucleosynthesis, existence of minima and fifth-force constraints.  Our results show that 
it is very difficult to find a consistent parameter range to satisfy two or more regimes.

The tightest constraint seems to come from BBN.  If the curvature values are known today and during the BBN 
epoch, the parameter $b$ would be constrained to a high degree.  Comparison of this result with fifth force 
constraints can ultimately decide the viability of this theory. Of course, more complicated cases for $f(R)$ 
might be considered, which might evade the afore--mentioned problems, see e.g. the discussion in \cite{Nojiri:2006gh}.

In future, one should also verify our results by solving the full field equations in the presence of an extended object, like the Earth, and study the stability of the solutions, similar to \cite{Faraoni:2005ie,Cognola:2005sg}.

\section*{Acknowledgements}
The authors wish to acknowledge S.~Nojiri and S.D.~Odintsov for early communications. We are grateful to 
A.~de Felice, S.~Nojiri and S.D.~Odintsov for comments on an earlier draft of the paper. We also wish to thank 
D.~Easson and K.~van~Acoleyen for ongoing discussions. LMHH and AWB acknowledge support from PPARC.

\bibliographystyle{h-physrev3}
\bibliography{paper}

\end{document}